\documentclass[pdflatex,sn-nature]{sn-jnl}

\usepackage{graphicx}%
\usepackage{multirow}%
\usepackage{amsmath,amssymb,amsfonts}%
\usepackage{amsthm}%
\usepackage{mathrsfs}%
\usepackage[title]{appendix}%
\usepackage{xcolor}%
\usepackage{textcomp}%
\usepackage{manyfoot}%
\usepackage{booktabs}%
\usepackage{algorithm}%
\usepackage{algorithmicx}%
\usepackage{algpseudocode}%
\usepackage{listings}%
\usepackage{amsmath}%
\usepackage{amssymb}%
\usepackage{siunitx}%
\usepackage{bm}
\usepackage{xcolor}
\usepackage{hyperref}
\definecolor{lightblue}{RGB}{80,130,220}
\definecolor{dngreen}{RGB}{120,190,120}
\definecolor{afcolor}{HTML}{b3443c}

\def\gsim{\lower.5ex\hbox{\gtsima}} 
\def\lsim{\lower.5ex\hbox{\ltsima}} 
\def\gtsima{$\; \buildrel > \over \sim \;$} 
\def\ltsima{$\; \buildrel < \over \sim \;$} \def\gsim{\lower.5ex\hbox{\gtsima}} 
\def\lsim{\lower.5ex\hbox{\ltsima}} 
\def\simgt{\lower.5ex\hbox{\gtsima}} 
\def\simlt{\lower.5ex\hbox{\ltsima}}

\newcommand{\Msun}{\,{\rm M_\odot}}

\newcommand{\Mblack}{M_\bullet}

\newcommand{\kms}{{\rm km \, s^{-1}}}

\begin{document}

\title[The Little Red Dots Are Direct Collapse Black Holes]{The Little Red Dots Are Direct Collapse Black Holes}


\author*[1,2]{\fnm{Fabio} \sur{Pacucci}}\email{fabio.pacucci@cfa.harvard.edu}

\author[3]{\fnm{Andrea} \sur{Ferrara}}\email{andrea.ferrara@sns.it}

\author[4]{\fnm{Dale D.} \sur{Kocevski}}\email{dkocevsk@colby.edu}

\affil*[1]{\orgname{Center for Astrophysics $\vert$ Harvard \& Smithsonian}, \city{Cambridge}, \postcode{02138}, \state{MA}, \country{USA}}

\affil[2]{\orgdiv{Black Hole Initiative}, \orgname{Harvard University}, \city{Cambridge}, \postcode{02138}, \state{MA}, \country{USA}}

\affil[3]{\orgname{Scuola Normale Superiore}, \city{Pisa}, \postcode{50156}, \country{Italy}}

\affil[4]{\orgdiv{Department of Physics and Astronomy}, \orgname{Colby College}, \city{Waterville}, \postcode{04901}, \state{ME}, \country{USA}}


\abstract{\textbf{The discovery by \textit{JWST} of a substantial population of compact ``Little Red Dots'' (LRDs) presents a major puzzle: their observed spectra defy standard astrophysical interpretations. Here, we show that LRD spectra are naturally reproduced by emission from an accreting Direct Collapse Black Hole (DCBH). Using radiation-hydrodynamic simulations, we follow the growth of the DCBH seed via a dense, compressionally heated, collisionally ionized accretion flow. The model self-consistently reproduces the screen responsible for the observed Balmer absorption, while allowing UV/optical emission to partially escape, along with reprocessed infrared radiation. Crucially, this structure is not a blackbody and requires no stellar contribution: the UV continuum originates entirely from reprocessed DCBH radiation, attenuated only by a small amount of dust with an extinction curve consistent with high-redshift galaxies. This single framework simultaneously explains the key observational puzzles of LRDs: (a) weak X-ray emission, (b) metal and high-ionization lines alongside absent star-formation features, (c) overmassive black holes, (d) compact morphology, (e) abundance and redshift evolution -- linking them directly to pristine atomic-cooling halos, (f) long-lived ($>100$ Myr), slowly variable phases driven by radiation pressure. Our findings indicate that \textit{JWST} is witnessing the widespread formation of heavy black hole seeds in the early Universe.}}


\maketitle

The James Webb Space Telescope (\textit{JWST}) was designed to explore the distant Universe. 
Among other discoveries, \textit{JWST} revealed a new class of astrophysical sources: the ``Little Red Dots'' (LRDs, \cite{Matthee_2023, Kocevski_2023, Harikane_2023}), which have become the most investigated mystery in early cosmic evolution.

Initially interpreted as massive star-forming galaxies \cite{Labbe_2023}, their tiny effective radii of $<100$ pc \cite{Baggen_2023, Furtak_2023} challenged this hypothesis \cite{Guia_2024, Pacucci_Hernquist_2025}. If solely made of stars, LRDs would represent an unexplored regime of galaxy formation in the $\Lambda$CDM framework \cite{BFPR_1984}.
An alternative class of models suggests that a massive black hole (MBH) powers the LRDs. This hypothesis has caveats, as the MBHs would be (i) overmassive relative to their host stellar mass \cite{Pacucci_2023_JWST, Maiolino_2023_new} and (ii) undetected in the X-ray \cite{Pacucci_Narayan_2024, Maiolino_2024_Xray, Lambrides_2024}.

Additional peculiarities of the LRDs include a V-shaped spectral energy distribution (SED) with a strong Balmer break \cite{Setton_2024} and a spectral shape at longer wavelengths similar to a blackbody \cite{DeGraaff_2025}. Their spectra also feature high-ionization forbidden lines -- typically observed around accreting MBHs -- but a lack of equivalently strong star-formation lines \cite{Lambrides_2025}. Interestingly, some high-ionization lines, such as [NeV], are absent or very weak.

This study demonstrates that all observed properties of LRDs are explained by identifying them with accreting Direct Collapse Black Holes (DCBHs). DCBHs were proposed as a channel to form heavy black holes ($\Mblack \sim 10^5 \Msun$) at high redshifts \cite{Loeb_Rasio_1994, Bromm_Loeb_2003, Lodato_Natarajan_2006, Ferrara_2014}, seeding the growth of supermassive black holes in quasars ($\Mblack \gtrsim 10^9 \Msun$ at $z > 6$ \cite{Fan_2001, Mortlock_2011, Wu_2015}). 

\section{Direct Collapse Black Hole Model}\label{sec:DCBH}
This study is based on a suite of radiation-hydrodynamic (RHD) simulations developed to model the emission properties of DCBHs \cite{Pacucci_2015, PFVD_2015}. The simulations follow time-dependent, nearly spherical gas accretion onto a black hole seed of initial mass $\Mblack = 10^5\,\Msun$, at rest at the center of an atomic-cooling halo of total mass $M_h = 6.2 \times 10^7 \Msun$ at redshift $z=10$. We self-consistently solve for gas inflow, radiation pressure, and heating/cooling processes. 

\begin{figure}[h]
\centering
\includegraphics[width=1.0\textwidth]{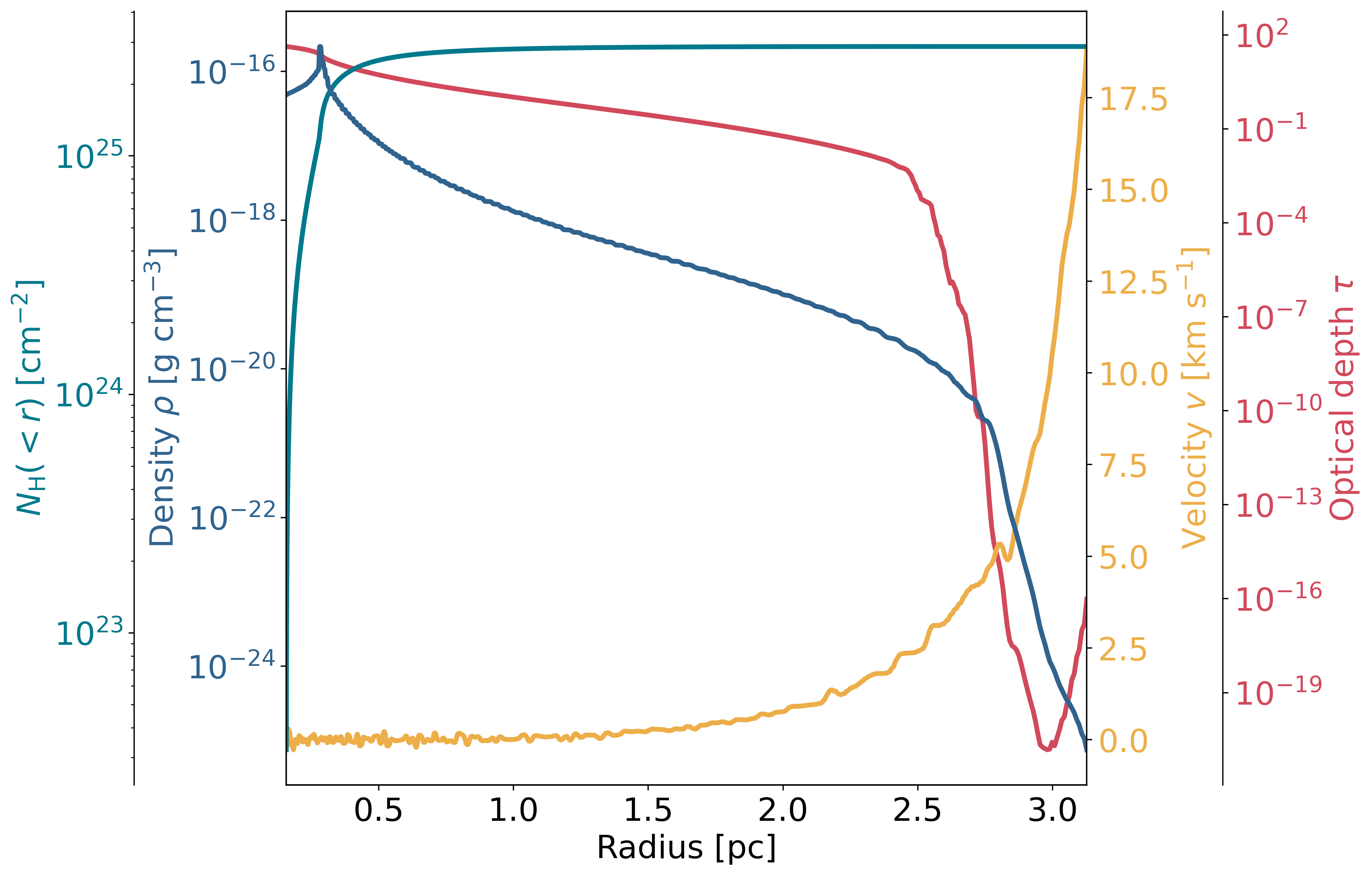}
\caption{Radial profiles of gas density, cumulative hydrogen column density, velocity, and optical depth at $75$ Myr after DCBH seeding, corresponding to a representative, intermediate evolutionary stage of the RHD simulation \cite{Pacucci_2015}. These profiles are used to generate the DCBH spectrum in Fig. \ref{fig:spectrum}.}
\label{fig:profiles}
\end{figure}

\begin{figure}[h]
\centering
\includegraphics[width=1.0\textwidth]{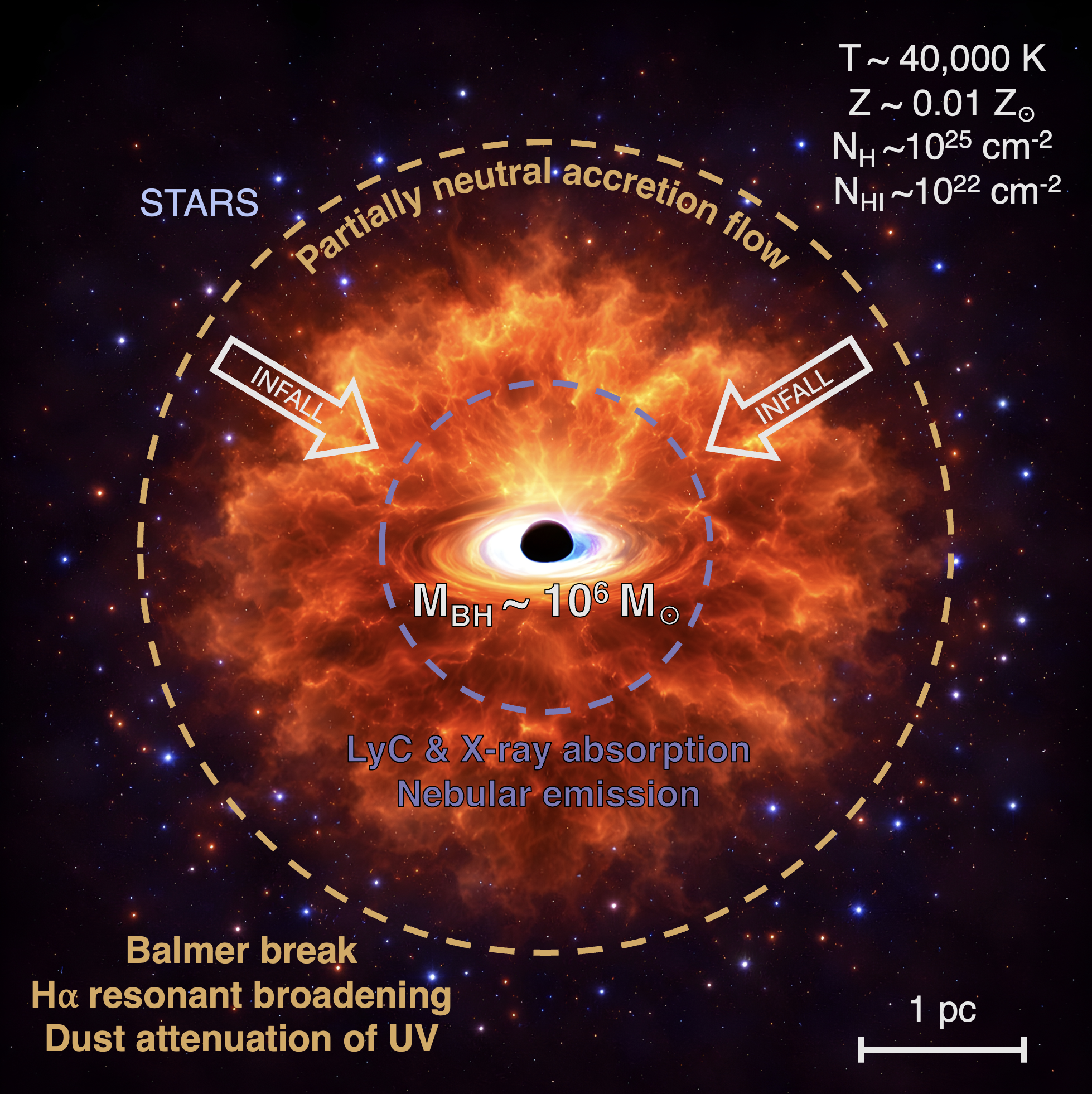}
\caption{Artist's illustration of our DCBH model for Little Red Dots.}
\label{fig:illustration}
\end{figure}

Figure~\ref{fig:profiles} displays typical radial profiles for gas density $\rho$, cumulative column density $N_{\rm H}$, velocity $v$, and optical depth $\tau$. The RHD simulations compute the optical depth for the frequency-integrated radiation field, and include electron scattering, bound-free absorption, and H$^{-}$ opacity in a metal-free gas \cite{Pacucci_2015}. These profiles are shown $75$ Myr after DCBH seeding, corresponding to an intermediate evolutionary stage between seed formation and the depletion of the available gas \cite{Pacucci_2015}. Figure \ref{fig:illustration} displays an artist's illustration of our DCBH model for LRDs.

The inner regions are characterized by high densities reaching $>10^{7-8} \, \rm cm^{-3}$ in the inner core, and high temperatures due to compressional heating, caused by accretion onto the DCBH \cite{Pacucci_2015}. 
As a result of the process leading to the DCBH formation, the density distribution of the halo gas (with mass $M_g \approx 10^7 \Msun$) has contracted from the virial radius of the parent halo ($\approx 1 $ kpc) to a few parsecs \cite{Schleicher_2013}. At $75$ Myr after seeding (i.e., the RHD snapshot used in this study), the total cumulative gas column density has increased to $N_{\rm H}\approx 3\times 10^{25} \rm cm^{-2}$; thus, the system is heavily Compton-thick.

The frequency-integrated optical depth is $\tau \gg 1$, thus the core is optically thick up to a radius of $\sim 1$ pc; outside, the optical depth decreases rapidly. The radiation from the DCBH accretion disk does not significantly affect the accretion flow temperature or ionization state. Most of the gas settles to a temperature of $T \approx 4\times10^4$ K due to a balance between compressional heating and radiative cooling (see Supplementary Materials, Sec. \ref{suppl:th_balance}). The accretion flow is, thus, almost completely ionized (neutral fraction $\approx 10^{-3}$); the relic neutral medium produces the observed Balmer absorption, described in Sec. \ref{sec:LRD_spectrum}.

The internal region is nearly hydrostatic ($v \approx 0$), while the most external layers are outflowing at the relatively low speed of $\sim 10-20 \, \rm km \, s^{-1}$ (see Sec. \ref{sec:inferred_properties}). Accretion proceeds intermittently due to radiative feedback, featuring long-lived, mildly super-Eddington (on average $L = 1.35 \,  L_{\rm Edd}$) accretion phases \cite{Pacucci_2015}.

\section{Synthetic LRD Spectra}\label{sec:LRD_spectrum}
To compute the observed spectrum, the outputs from the RHD simulations are post-processed with \textsc{Cloudy} \cite{Cloudy}. Full frequency-dependent radiative transfer through the accretion flow is performed using the simulated gas density profile and a physically-motivated input spectrum \cite{Yue_2013}, self-consistently accounting for absorption, reprocessing, and nebular emission (see Supplementary Materials, Sec. \ref{suppl:Cloudy}). This framework has been used to predict the observable signatures of heavy black hole seeds \cite{PFVD_2015}, identify the first candidate DCBHs \cite{Pacucci_2016_DCBH, Pacucci_2017_CR7}, interpret \textit{JWST} observations \cite{Natarajan_2023, Nabizadeh_2024}, and assess prospects for next-generation X-ray observatories \cite{Pacucci_2019BAAS, AXIS_2023}.

The key spectral features can be understood as follows. The input DCBH accretion disk spectrum, extending to energies $\simgt 300\ $keV, is filtered by the Compton-thick accretion flow before escaping. As a result, photons in the energy range $13.6\ {\rm eV} < E < 1\ {\rm keV}$ suffer complete photoelectric absorption; more energetic X-rays can instead (partially) escape. Hence, these objects contribute negligibly to cosmic reionization. The absorbed energy is re-emitted as ``nebular emission'' with contributions from free-free, free-bound, and two-photon processes. Ly$\alpha$ emission is not included, as at these large column densities, Ly$\alpha$ photons are resonantly trapped in the surrounding gas \citep{Ferrara25b}. Hence, $n = 2 \to n = 1$ transitions will eventually produce two-photon emission. 

The nebular emission boosts the spectrum below $13.6$ eV. However, as a sizable fraction of H-atoms is in the $n=2$ state at any time, photoelectric absorption from the $n=2$ level selectively suppresses the flux at $\lambda<\lambda_{\rm B} =3646$~\AA, converting the intrinsic Balmer jump produced by the free-bound continuum into the observed Balmer break, which is naturally reproduced in our setup.  

Finally, we account for the presence of dust, which explains the observed reddening of the LRD spectrum, especially in the UV. As the gas temperature settles to $T \approx 4\times10^4$ K, grain sputtering is inefficient, and dust survives without issues, at least in the external layers where the density is sufficiently low ($n \simlt 10^8\ \rm cm^{-3}$) that gas–dust thermal coupling is inefficient. We find that a very small dust amount ($12.9 \Msun$) is required to fit the data (see Supplementary Materials, Sec. \ref{suppl:dust_attenuation}). We adopt a dust attenuation law, $A_\lambda$, that accurately reproduces the properties of \textit{JWST} galaxies at $z\sim2-12$ \cite{Markov_2025}. Dust attenuation is applied multiplicatively to the \textit{filtered} DCBH spectrum: $F_\lambda^{\rm obs} =  10^{-0.4 A_\lambda} F_\lambda^{\rm DCBH}$ to obtain the observed spectrum, $F_\lambda^{\rm obs}$.

\section{Comparison With JWST LRD Spectrum}
\label{sec:data_comparison}
To validate our DCBH-based model for LRDs, we compare it with recent \textit{JWST} data. We focus on the prototypical case of RUBIES-EGS~42046 at $z=5.28$, i.e., at the peak of the LRD redshift distribution \cite{Kocevski_2024, Pacucci_Loeb_2025}, and whose spectrum exhibits the defining characteristics of the class. Hence, we compare the $0.9$--$5.2\,\mu$m observed-frame spectrum of RUBIES-EGS~42046 to the synthetic spectrum predicted by our DCBH model for an initial seed mass of $10^5 \Msun$ \cite{Pacucci_2015, PFVD_2015}.
To determine the dust attenuation parameters implied by the data, we developed a Markov chain Monte Carlo (MCMC) fitting pipeline that compares the observed spectrum of RUBIES-EGS~42046 to our synthetic spectrum (see Supplementary Materials, Sec. \ref{suppl:mcmc}).

\begin{figure}[h]
\centering
\includegraphics[width=1.0\textwidth]{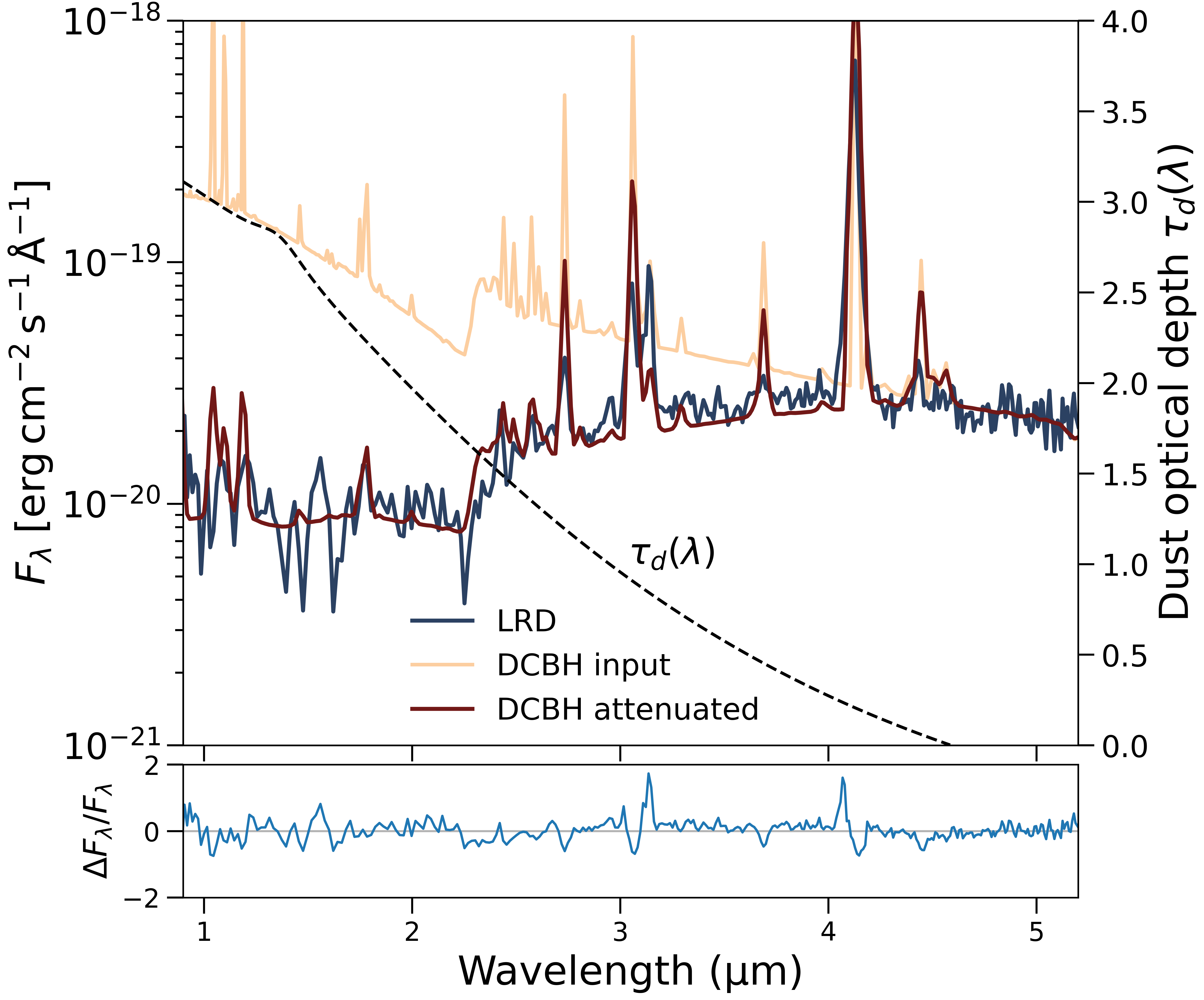}
\caption{Comparison between the attenuated DCBH spectrum (red) and the spectrum of a prototypical LRD at $z = 5.28$ (blue), RUBIES-EGS-42046 \cite{Kocevski_2024}. The unattenuated DCBH spectrum is also shown (orange), together with the dust attenuation law $\tau_d(\lambda)$. The bottom panel displays the residuals between the attenuated DCBH spectrum and the LRD spectrum. The DCBH spectrum is shown at the intermediate accretion stage of $\sim 75$ Myr, when it reached a mass of $\sim 10^6 \Msun$ \cite{Pacucci_2015}.}
\label{fig:spectrum}
\end{figure}

The predicted DCBH spectrum is shown in Fig. \ref{fig:spectrum}, where it is compared to RUBIES-EGS~42046. The DCBH spectrum reproduces the LRD's spectrum self-consistently, including both the blue and red V-arms, without any need for arbitrary stellar components. The residuals from our spectral fitting procedure are typically $<10\%$, indicating an excellent fit. The only spectral regions where they become larger are in correspondence with specific emission lines, such as the [OIII]$\lambda4959+\lambda5007$ doublet. The large $\rm H\alpha/H\beta \simeq 10$ ratio is due to dust, consistent with observations. We find a moderate rest-frame visual attenuation of $A_V \simeq 0.7$~mag; this relatively low value alleviates the long-standing concern that the large extinctions often inferred for LRDs would inevitably imply strong hot dust re-emission at long wavelengths (see Sec.  \ref{sec:inferred_properties}).

\section{Inferred LRD Properties}
\label{sec:inferred_properties}
As our model successfully reproduces the spectra of LRDs, we can now examine their inferred physical properties. 
\vspace{0.3cm}

\noindent\textbf{Stellar mass} -- If supernovae (SNe) are the main contributors to the dust mass required by the attenuation ($12.9 \Msun$, Sec. \ref{sec:LRD_spectrum}), the associated stellar mass formed in the LRD is  
\begin{equation}
    M_\star = M_d (y_d \nu)^{-1} = 6824\ \Msun \, ,
\end{equation}
where $y_d = 0.1 \Msun$ and $\nu^{-1} = 52.9 \Msun$  are the net dust yield per SN, and the number of SNe per unit stellar mass formed, respectively \citep{Ferrara25}. Hence, $N_{\rm SN}=129$ supernovae are required, corresponding to $M_\star=6824 \Msun$. 

\vspace{0.3cm}

\noindent\textbf{Metallicity} -- Using similar arguments for metals, assuming a SN metal yield $y_Z = 2.44 \, \Msun$ \citep{Ferrara25}, we find that a metal mass $M_Z = 315 \Msun$ was injected in the remaining fraction ($\approx 50\%$) of the initial halo gas mass ($9.6\times 10^6 \Msun$) that was not yet accreted by the DCBH after $75$ Myr. This corresponds to an average gas metallicity $Z = M_Z/M_g \simeq 5\times 10^{-3} \, \rm Z_\odot$. Such non-zero metallicity is consistent with recent studies \cite{Maiolino_2025}, accounts for the presence of some metal lines in the observed spectrum, and is used as a reference value in our \textsc{Cloudy} simulations.

\vspace{0.3cm}

\noindent\textbf{Overmassive black hole} -- After $75$ Myr, the DCBH has grown to $M_\bullet \approx 3\times10^6\ \Msun$ \cite{Pacucci_2015}. This implies a black hole-to-stellar mass ratio of $M_\bullet/M_\star \sim 400$, i.e., the DCBH is overmassive, as predicted by detailed simulations of heavy seeds \cite{Scoggins_2023}.

\vspace{0.3cm}

\noindent \textbf{Broad line components} -- The H$\alpha$ line in the \textit{JWST}/NIRSpec PRISM spectrum of RUBIES-EGS~42046 has a measured FWHM of $3300 \pm 60\ {\rm km\ s^{-1}}$ \citep{Jones_2025}. 
Because PRISM has a low and strongly wavelength-dependent resolving power, $R(\lambda)\sim 30-300$ \citep{NIRSPEC_2022}, unresolved spectral features are broadened by the instrumental line-spread function, corresponding to a velocity resolution floor $\Delta v \sim c/R \sim 3000 \, {\rm km\ s^{-1}}$ across the bandpass.
For this reason, line widths measured from PRISM data alone generally reflect a combination of intrinsic broadening and instrumental smearing.

In our spectral modeling \cite{PFVD_2015}, nebular emission is computed with \textsc{Cloudy}; in the absence of turbulence, lines are characterized by a negligible width and the intrinsic line profiles are narrow compared to the PRISM resolution (see Fig. \ref{fig:spectrum}). 
To compare the model to the observed SED, we convolve the attenuated model spectrum with a wavelength-dependent approximation to the PRISM line-spread function (see Supplementary Materials, Sec.~\ref{suppl:projection}), thus enforcing the finite instrumental resolution on the model. 

In general, the broad components might arise from a broad-line region associated with virialized gas in the vicinity of the DCBH. For virial motions, $v \simeq (GM_\bullet/R)^{1/2}$, so ${\rm FWHM}\sim 3000\ \mathrm{km\,s^{-1}}$ requires emission from radii $R \sim 10^{-3}\ \mathrm{pc}$ ($\sim$ a few light-days) for $M_\bullet\sim 3\times 10^{6} \Msun$. If instead the broad-line region size follows the empirical radius-luminosity relation for AGN, the expected virial widths are substantially smaller than the observed values.
In addition to gravitational broadening, broad Balmer wings can be produced by resonant scattering in the dense, extended gas distribution. In this scenario, line photons produced in the H\,\textsc{ii} region undergo multiple scatterings off neutral or partially ionized hydrogen before escaping, leading to substantial frequency diffusion and the formation of broad wings \cite{DeGraaff_2025, Rusakov_2025, Chang_2025, Naidu_2025a, Torralba_2025b}.
\textsc{Cloudy} does not include radiative transfer effects associated with resonant scattering; hence, the widths of the Balmer lines produced by the code are to be interpreted as lower limits.

\vspace{0.3cm}

\noindent\textbf{Dust continuum emission} -- Dust in the gas surrounding the DCBH and out to the external optically thin gas layer ($\approx 1$ pc, see Fig. \ref{fig:profiles}) absorbs the UV/optical radiation and re-emits it in the infrared. We derive a dust temperature $T_d = 254$ K (see Supplementary Materials, Sec. \ref{suppl:dust_emission}), corresponding to a peak observed-frame emission at $\lambda_p = 72.2 \, \rm \mu m$. The observed flux is $F_{\lambda_p}^{\rm obs}
= 78.4$ nJy. This level of far-IR emission is far below the sensitivity of current facilities; moreover, the predicted Rayleigh-Jeans tail at longer wavelengths remains consistent with upper limits from deep sub-mm observations \cite{Casey_2024}.

\section{Implications of the DCBH Nature of the LRDs}\label{sec:implications}
Interpreting the LRDs as DCBHs has several physical consequences. Below, we outline some immediate implications regarding their evolutionary stage, variability, multi-wavelength appearance, and formation pathways. For completeness, we assess whether our DCBH interpretation is consistent with the full known set of LRDs observational constraints \cite{Inayoshi_2025_review}.

\subsection{Intrinsic X-ray Weakness}
A defining observational property of LRDs is their extreme X-ray weakness \cite{Maiolino_2024_Xray, Yue_2024_Xray, Pacucci_Narayan_2024}. This behavior arises naturally in the DCBH scenario, where the black hole is embedded in a Compton-thick ($N_{\rm H} \gtrsim 1.5 \times 10^{24} \, \rm cm^{-2}$) gas distribution for a significant fraction of its active lifetime of $>100$ Myr \cite{Pacucci_2015, PFVD_2015}. As displayed in Fig. \ref{fig:profiles}, the spectrum used in this study is generated within an accretion flow with a column density $N_{\rm H} \approx 3 \times 10^{25} \, \rm cm^{-2}$.
Studies from a decade ago \cite{Pacucci_2015, PFVD_2015} indicated that such systems should be heavily obscured in X-rays, with high-energy photons efficiently absorbed and reprocessed to lower energies (e.g., Fig. 3 of \cite{PFVD_2015}). Only during the final stages, when the gas column density rapidly drops, does the X-ray emission escape.
Because DCBH accretion is characterized by mildly super-Eddington phases, recent studies \cite{Pacucci_Narayan_2024, Madau_2024} suggest that the intrinsic X-ray emission may be even weaker than originally predicted. The full DCBH spectrum used in this study is displayed in Fig. \ref{fig:DCBH_full_spectrum} of the Supplementary Materials, with additional details on the X-ray weakness.

\subsection{Long-lived Accretion Phases and Outflow Profiles}
This study focuses on a DCBH at an evolutionary time of $\sim 75$ Myr after seed formation. RHD simulations of DCBHs \cite{Pacucci_2015} show that this epoch corresponds to the central stages of the accretion history, when $\sim 50\%$ of the available gas has been consumed, and the system is still far from its terminal ``last-gasp'' phase. In these simulations, accretion proceeds for $\sim 100-200$ Myr, characterized by mildly super-Eddington phases \cite{Pacucci_2015}, before a final radiation-driven event expels the remaining gas from the host. 
Signatures of this transition include the emergence of outflows, which are explicitly seen as positive gas velocities in the RHD simulations, as displayed in Fig. \ref{fig:profiles}. Such outflows provide a physical framework for the presence of P-Cygni-like features and asymmetric Balmer-line profiles observed in LRDs \cite{Torralba_2025b}.

\subsection{Late-Stage DCBH Evolution}
The late stages of DCBH accretion may be associated with observational properties that are now being identified in a subset of LRDs and related populations. For instance, \cite{Hviding_2026} report the detection of an X-ray-luminous LRD at $z=3.28$. This source may correspond to a late DCBH evolutionary stage, close to the ``last gasp'' \cite{Pacucci_2015}, in which a substantial fraction of the surrounding gas has already been depleted, resulting in a Compton-thin absorbing column, as originally pointed out in \cite{PFVD_2015}.
Additional support for this evolutionary interpretation comes from the identification of Little Blue Dots (LBDs, \cite{Asada_2026}). These systems share several structural and spectroscopic similarities with LRDs, but lack the red optical continua that characterize them. Within a DCBH framework, LBDs may represent a transitional phase in which the system has largely emerged from its dust-enshrouded state, revealing a bluer continuum and AGN-like spectral properties.

\subsection{Intrinsic Variability Driven by Radiation Feedback}
Most LRDs show weak or no variability, while a small fraction exhibits measurable changes \cite{Secunda_2026, Zhang_2025}. 
The RHD simulations predict intrinsic, self-regulated variability as an inevitable consequence of the coupling between accretion and radiative feedback \cite{Pacucci_2015}. As radiation pressure intermittently halts and restarts the inflow, the emitted luminosity fluctuates on timescales set by the dynamical response of the gas. This variability arises from the self-regulated nature of accretion in optically thick environments and provides a natural explanation for the low-level, long-timescale variability observed in LRDs \cite{Zhang_2025}. Importantly, this intrinsic, low-amplitude variability is expected to be strongest toward the end of the accretion history, when the gas reservoir becomes increasingly susceptible to radiative disruption \cite{Pacucci_2015}.

\subsection{Continuum Shape and Balmer-series Features}
The defining spectral properties of the LRDs are naturally reproduced within our framework. The V-shaped spectrum, with a turnover at rest-frame $\sim 3000 - 4000$~\AA, emerges from the combined effects of reprocessed UV/optical radiation emerging from the accretion flow and some modest dust attenuation. The location of the turnover is set primarily by hydrogen bound-free opacity and reprocessing in the outer layers, rather than by a fine-tuned combination of stellar and AGN components, explaining the remarkable uniformity of LRD continua.

Within the same geometry, Balmer emission lines are produced in the ionized region surrounding the DCBH. As Balmer photons propagate through the extended, optically thick layer, scattering and radiative transfer generate broad wings. The coexistence of broad emission with Balmer absorption and a pronounced Balmer break further suggests the presence of neutral medium within the partially-ionized, high-density accretion flow, whose thermodynamical state is predicted by the combination of our RHD simulations and \textsc{Cloudy}. Bound-free absorption by excited ($n=2$) neutral hydrogen in this layer creates the observed Balmer break, and also affects the Balmer decrement. 

\subsection{Compact Morphology}
The LRDs are unresolved or marginally resolved by \textit{JWST}, with inferred sizes $\lesssim 100$~pc \cite{Baggen_2023}. In the DCBH scenario, the dominant emission originates from the nuclear region and the surrounding gas on parsec scales, naturally producing point-like sources. Any extended emission is expected to be energetically subdominant, consistent with observations.

\subsection{Abundance and Redshift Evolution}
The observed abundance and redshift evolution of LRDs provide an essential consistency check of the DCBH interpretation. The comoving number density of LRDs \cite{Kocevski_2024} is comparable to the abundance of heavy seeds predicted by DCBH models \cite{Jeon_2025}. Light seed models that rely on super-Eddington accretion tend to overproduce the observed population, while heavy seed scenarios naturally yield rarer objects. Recent semi-analytical modeling demonstrates that DCBHs reproduce both the observed LRD number densities and their host halo properties more successfully than light seeds \cite{Jeon_2025}.

Equally compelling is the redshift dependence of the LRD population ($4 \lesssim z \lesssim 9$, \cite{Kocevski_2024}). DCBHs are expected to form in pristine, atomic-cooling halos, whose abundance declines rapidly as structure growth progresses \cite{Loeb_Rasio_1994,Bromm_Loeb_2003, Lodato_Natarajan_2006}. As the Universe evolves, the conditions required for direct collapse become increasingly rare \cite{Jeon_2025}. The observed decline of the LRD population with cosmic time closely mirrors this theoretical expectation, providing a natural explanation for why such objects are predominantly detected at $z \gtrsim 4$ \cite{Pacucci_Loeb_2025}. In this framework, the emergence and disappearance of the LRDs trace the cosmological window during which atomic-cooling halos can host the birth and the subsequent growth of heavy black hole seeds.
Additionally, the formation of a DCBH requires a nearby source of Lyman-Werner radiation \cite{DB_1996, Visbal_2014b}, a condition that is consistent with observational evidence indicating that LRDs are preferentially clustered \cite{Tanaka_2024}.

\vspace{0.3cm}
Taken together, these considerations show that the DCBH model provides a self-consistent explanation for the observational properties of LRDs. Our results strongly suggest that \textit{JWST} is witnessing the widespread formation of heavy black hole seeds in the early Universe.

\newpage

\backmatter

\section*{\LARGE Supplementary Materials}
The Supplementary Materials provide additional details on the models used and a broader perspective on some topics discussed in the main text.

\setcounter{section}{0}

\section{Cosmological Model}\label{suppl:cosmo_model}
Throughout this study, and consistent with the original papers that developed the DCBH model \cite{Pacucci_2015, PVF_2015, PFVD_2015, Pacucci_2017_CR7}, we assume a flat Universe with the following cosmological parameters: $\Omega_{\rm M} = 0.32$, $\Omega_{\Lambda} = 1- \Omega_{\rm M}$, $\Omega_{b} = 0.05$, $h=0.67$, $n_s = 0.96$, $\sigma_8=0.83$. Here, $\Omega_{\rm M}$, $\Omega_{\Lambda}$, and $\Omega_{b}$ are the total matter, vacuum, and baryon densities, in units of the critical density; $h$ is the Hubble constant in units of $100 \, \kms \rm Mpc^{-1}$, $n_s$ is the primordial matter power spectrum index, and $\sigma_8$ is the late-time fluctuation amplitude parameter \citep{Planck14}.

\section{CLOUDY Photoionization Modeling}\label{suppl:Cloudy}
This Section details how the photoionization code \textsc{Cloudy} (version C25, \cite{Cloudy}) was used to generate the observed spectrum described in Sec. \ref{sec:LRD_spectrum}.
We adopted a spherically symmetric gas distribution and an input AGN spectrum for the ionizing continuum \cite{Yue_2013}; this setup follows the same configuration originally adopted in \cite{Pacucci_2015, PFVD_2015}.

\paragraph{Geometry:} We model the emitting/absorbing medium as a spherical shell extending from an inner radius
$\log_{10}(R_{\rm in}/{\rm cm}) = 17.69$ to an outer (stopping) radius $\log_{10}(R_{\rm out}/{\rm cm}) = 18.97$, consistent with the extent of the density profile derived from the RHD simulations \cite{Pacucci_2015}.
This configuration is intended to represent the circumnuclear gas in the RHD simulations at the selected evolutionary stage of $75$ Myr, midway between seeding and gas depletion.

\paragraph{Input ionizing continuum:}
The incident continuum is provided via an explicit tabulation of $\log \nu$ versus $\log F_\nu$. This table encodes a standard AGN-like spectrum appropriate for a $\sim 10^{5-6} \Msun$ black hole \cite{Yue_2013}. In that model, the intrinsic source spectrum can be described as the sum of three components: (i) a multi-color blackbody, (ii) a power-law, and (iii) a reflection component, and it extends from the far-infrared to hard X-rays (up to $\sim$MeV energies).
In practice, \textsc{CLOUDY} linearly interpolates the tabulated values to construct the incident spectrum on its internal frequency grid.

The overall normalization of the ionizing source is set corresponding to a bolometric luminosity $L_{\rm bol} = 10^{45}\ {\rm erg\ s^{-1}}$ integrated over the entire input continuum, consistent with the original DCBH model \cite{Pacucci_2015}. We forced the code to iterate to convergence to ensure that the thermal and ionization solutions are self-consistent.

The thermal solution is computed internally by Cloudy, while we impose a minimum gas temperature to prevent the equilibrium solution from cooling below $4\times 10^4$ K. This temperature results from a balance between compressional heating and radiative cooling (see Supplementary Materials, Sec. \ref{suppl:th_balance}).

\paragraph{Radial density law:}
The hydrogen number density profile is specified using a dlaw table as a tabulated function of radius, thereby enforcing the RHD-derived density structure within the shell \citep{Pacucci_2015}.
Additional grid controls are used to regulate the spatial stepping and the total number of zones used to resolve gradients in the dense inner regions.

\paragraph{Chemical composition:}
We adopted a low-metallicity mixture with metals scaled to $10^{-2}$ of solar abundance, while disabling molecule formation. This metallicity level is compatible with our theoretical estimate in Sec. \ref{sec:inferred_properties}, derived from the dust attenuation law required to reproduce the observed spectrum.

\section{Full DCBH Spectrum and X-ray Weakness} \label{suppl:full}
\begin{figure}[ht]
\centering
\includegraphics[width=1.0\textwidth]{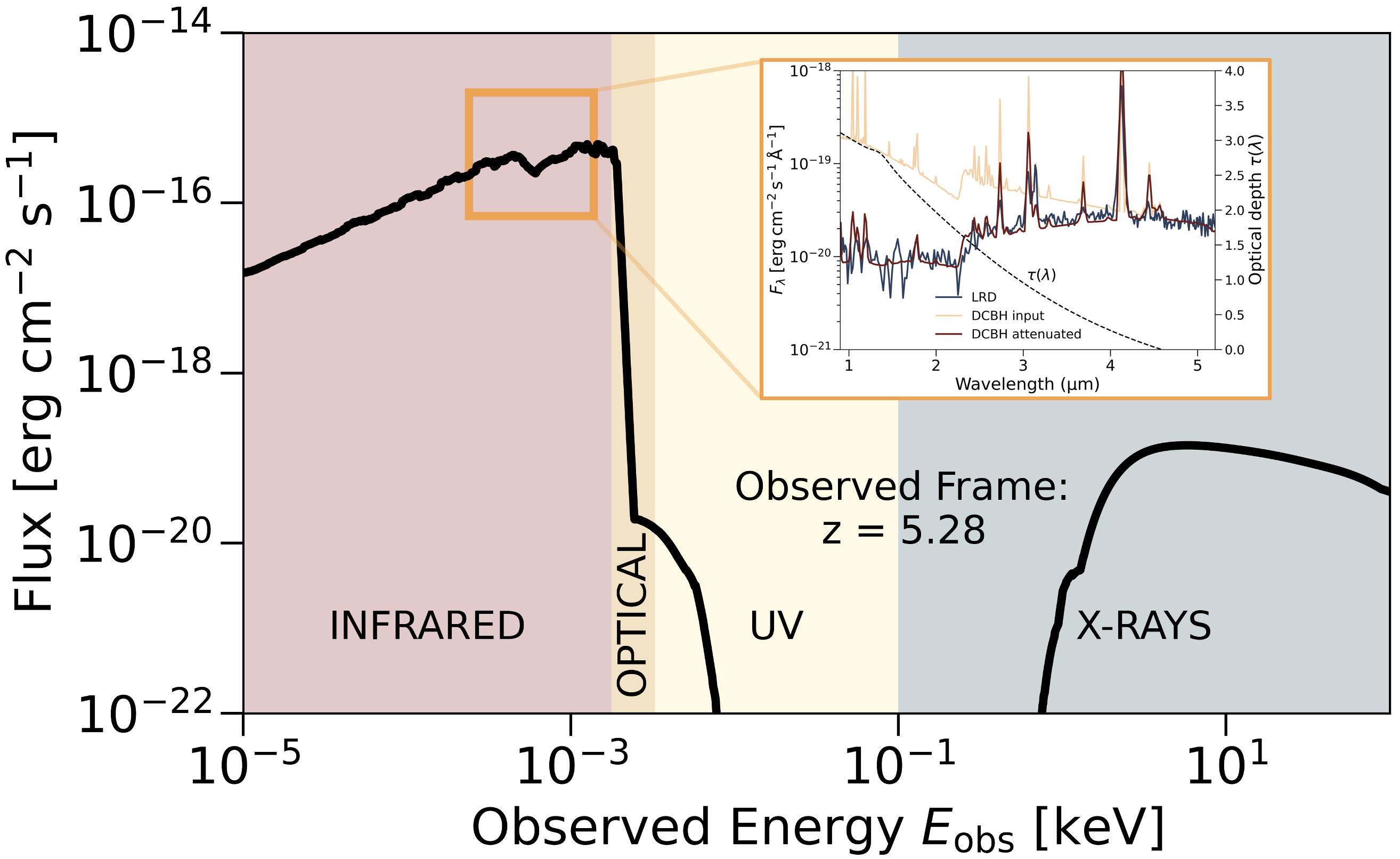}
\caption{Reference DCBH synthetic spectrum, from the far-infrared ($120 \, \rm \mu m$) to the hard X-rays ($100$ keV). Line smoothing was applied to improve the clarity of the continuum visualization. The inset shows a zoom-in of the wavelength range highlighted in Fig. \ref{fig:spectrum}. In this particular run, as the column density is $>10^{25} \, \rm cm^{-2}$, the X-ray emission is faint. This fact does not imply that all DCBH accretion phases are X-ray weak to this extent; on the contrary, in the final phases, when most of the gas is depleted, DCBH accretion is X-ray bright \cite{PFVD_2015}.}
\label{fig:DCBH_full_spectrum}
\end{figure}

For completeness, we display in Fig. \ref{fig:DCBH_full_spectrum} the full DCBH synthetic reference spectrum, from the far-infrared to the hard X-rays.

We caution the reader that the fact that this particular simulation is characterized by a weak X-ray emission of $\sim 10^{-19} \, \rm erg \, cm^{-2} \, s^{-1}$ does not imply that all DCBH accretion phases are X-ray weak to this extent. As shown in the original study developing the spectral analysis \cite{Pacucci_2015, PFVD_2015}, the observed X-ray emission at $\sim 1$ keV can reach $\sim 10^{-17} \, \rm erg \, cm^{-2} \, s^{-1}$ in the final phases ($\sim 150\ \rm Myr$) when most of the gas is depleted.

\section{Spectrum Projection Onto the JWST/PRISM Grid} \label{suppl:projection}
The intrinsic DCBH spectrum produced by \textsc{Cloudy} is defined on the code's native wavelength grid and does not incorporate the finite spectral resolution of \textit{JWST}/NIRSpec. 
A direct comparison to the observed PRISM data, therefore, requires modeling of instrumental effects on the model spectrum.

The NIRSpec PRISM mode has a low and strongly wavelength-dependent resolving power, $R(\lambda)\sim 30-300$, which corresponds to a minimum velocity resolution $\Delta v \sim c/R\sim 3000 \, {\rm km\ s^{-1}}$ across the $0.6$--$5.3\,\mu$m bandpass \citep{NIRSPEC_2022}. 
Any intrinsic spectral feature narrower than $\Delta\lambda \simeq \lambda/R$ will therefore appear broadened in the observed spectrum.

Our comparison pipeline consists of two steps. First, we evaluate the model on the observed wavelength grid using interpolation, which enables point-by-point residuals relative to the data. 
Second, we convolve the attenuated model spectrum with an approximation to the PRISM line-spread function \citep{NIRSPEC_2022}. 
This convolution applies only the finite instrumental resolution to the model, enabling a like-for-like comparison between the synthetic spectrum and PRISM data.

\section{Dust Attenuation}\label{suppl:dust_attenuation}
Section \ref{sec:LRD_spectrum} describes the necessity of some dust attenuation. As detailed in the main text, we use the attenuation curve formalism developed in \cite{Markov_2025} to match \textit{JWST} galaxies observed in the redshift range $z\sim2-12$, i.e., a cosmic time encompassing when the totality of LRDs is observed.
Following the analytical fit from \cite{Markov_2025}, the attenuation curve is expressed in wavelength space as:
\begin{equation}
\begin{aligned}
\frac{A_\lambda}{A_V} =\;&
\frac{c_1}{(\lambda/0.08)^{c_2} + (0.08/\lambda)^{c_2} + c_3}
\\
&+\,233\,
\frac{\left[1-\frac{c_1}{6.88^{c_2}+0.145^{c_2}+c_3}-\frac{c_4}{4.60}\right]}
{(\lambda/0.046)^2 + (0.046/\lambda)^2 + 90}
\\
&+\,\frac{c_4}{(\lambda/0.2175)^2 + (0.2175/\lambda)^2 - 1.95} \, ,
\end{aligned}
\label{eq:dust_markov}
\end{equation}
where $\lambda$ is in $\mu$m and the four dimensionless coefficients $(c_1,c_2,c_3,c_4)$ fully specify the shape of the attenuation curve. This formulation encompasses a wide range of attenuation laws, including Calzetti-, SMC-, and MW-like curves, while allowing for the flatter slopes and weaker UV bumps inferred in high-redshift \textit{JWST} galaxies.
Details of the derived dust attenuation parameters are provided in the following Sec. \ref{suppl:mcmc} of the Supplementary Materials.

\section{Spectral Fitting}\label{suppl:mcmc}
As described in Sec. \ref{sec:data_comparison} of the main text, we developed an MCMC fitting pipeline that compares the observed LRD continuum to our synthetic DCBH spectrum.

At each step, the intrinsic DCBH spectrum is reddened using the four-parameter attenuation curve $(c_1,c_2,c_3,c_4)$. The posterior probability is sampled assuming Gaussian uncertainties in flux space, with a likelihood defined by the $\chi^2$ statistic evaluated over the full observed-frame wavelength range. We adopt broad priors on all attenuation parameters and report the maximum posterior solution as the best fit, with optimal parameters listed in Table \ref{table:fit_parameters}. These values of the dust attenuation curve are broadly in agreement with those found by \cite{Markov_2025} for \textit{JWST} galaxies in the same redshift range as LRDs.

\begin{table}[h]
\centering
\caption{Dust attenuation parameters; MCMC modeling}
\label{table:fit_parameters}
\begin{tabular}{lc}
\hline
\textbf{Parameter} & \textbf{Best-fit value} \\
\hline
$c_1$ & $16.480$ \\
$c_2$ & $0.652$ \\
$c_3$ & $0.989$ \\
$c_4$ & $0.01117$ \\
\hline
\end{tabular}
\end{table}

\section{Dust Emission}\label{suppl:dust_emission}
Our model requires a small amount of dust to attenuate the intrinsic DCBH spectrum, as detailed in Sec. \ref{sec:LRD_spectrum}. It is thus warranted to check whether the emission from this small amount of dust should be detectable.

In our model, the dust optical depth at rest-frame $1500$ \AA\ is $\tau_{1500} \approx 3$, corresponding to $A_V \approx 0.67$ for our best-fit extinction curve. The total dust mass is   
\begin{equation}
    M_d = \frac{4\pi r_{\rm HI}^2}{\kappa_{1500}} \tau_{1500} = 4.3 \, \tau_{1500}\ \Msun, 
\end{equation}
where $\kappa_{1500}=1.26\times 10^5\ \rm cm^2 g^{-1}$ is the dust mass absorption at $1500$ \AA\ \citep{Ferrara24}, and $r_{\rm HI} = 3$ pc.
The dust will achieve a mean temperature given by \citep{Ferrara22}:

\begin{equation}\label{eq:Td}
    T_d = \left(\frac{L_{\rm UV}}{\Theta M_d}\right)^{1/(4+\beta_d)} = 305\ \tau_{1500}^{-1/6}\ \rm K \, ,
\end{equation}
where
\begin{equation}
    \Theta = \frac{8\pi}{c^2}\frac{\kappa_{158}}{\nu_{158}^{\beta_d}}\frac{k_B^{4+\beta_d}}{h_P^{3+\beta_d}}\zeta(4+\beta_d)\Gamma(4+\beta_d).
\end{equation}
The IR mass absorption coefficient, $\kappa_\nu = \kappa_{158}(\nu/\nu_{158})^{\beta_d}$ is canonically pivoted at wavelength $\lambda_{158} = c/\nu_{158} =158\ \mu$m. We take $\kappa_{158}$ and $\beta_d$ consistently with the adopted extinction curve, i.e., 
$\kappa_{158}=13.55\, {\rm cm}^2 {\rm g}^{-1}$,  and $\beta_d=2.0$; $\zeta$ and $\Gamma$ are the Zeta and Gamma functions, respectively. Other symbols have the usual meaning.  Hence, we find $\Theta = 5.33\times 10^{-6}$. 

Finally, we approximate the UV luminosity with the one predicted by our model at $1500$ \AA, i.e., $L_{\rm UV} \simeq L_{1500}$. 
From the (unattenuated) flux, $F_{\lambda}^{\rm obs}(1500 \, \rm \AA) \simeq 2\times 10^{-19} \, \rm erg\ s^{-1}\ cm^{-2}\ \AA^{-1}$ (see Fig. \ref{fig:spectrum}), we find
\begin{equation}
    L_{1500} = 1500\ {\rm \AA}\ \frac{4\pi d_L^2(z)}{(1+z)}F_{\lambda}^{\rm obs} = 1.5\times 10^{43}\ \rm erg\ s^{-1},
\end{equation}
where $d_L$ is the luminosity distance to  $z=5.28$ appropriate for RUBIES-EGS 42046. With these assumptions, from Eq. \ref{eq:Td} we obtain $T_d = 254$ K for $\tau_{1500} \simeq 3$. The dust emission peaks at a rest-frame wavelength $\lambda_p = 11.5\ \mu$m. 
The corresponding flux at $\lambda_p(1+z)=72.2\ \mu$m is 
\begin{equation}
F_{\lambda}^{\rm obs}
= \frac{1}{(1+z)}\,
\frac{L_{\lambda_{\rm p}}}{4\pi d_L^2}
= \frac{M_d\,\kappa_{\lambda_{p}}\,B_{\lambda_{p}}(T_d)}
{(1+z)\,d_L^2(z)}  = 78.4\ \rm  nJy.
\end{equation}

Due to the small dust mass involved ($12.9 \Msun$) and the resulting far-infrared peak at long wavelengths, the predicted dust continuum emission lies well below the sensitivity of current facilities. Moreover, the corresponding Rayleigh-Jeans tail at sub-mm wavelengths remains consistent with existing upper limits on the dust content of LRDs derived from deep observations \cite{Casey_2024}.

\section{Accretion Flow Thermal Balance}\label{suppl:th_balance}
Gas falling onto the central DCBH is compressionally heated and cooled by radiative processes, such as Ly$\alpha$, free-free, and recombination cooling. The actual temperature profile is obtained by balancing heating and cooling rates. These can be simply estimated if one assumes classical isothermal Bondi accretion. In that case, the heating rate is 
\begin{equation}
    \Gamma(x) = c_s^3 \frac{\rho_\infty y(x)}{r_B} \frac{1}{x^2}\frac{d}{dx}\left[x^2 y(x)\right] \, , 
\end{equation}
where $x=r/r_B$ if the radius normalized to the Bondi radius, $r_B$, of the DCBH; $y(x) = \lambda/\left[x^2 u(x)\right]$, where $\lambda = (1/4)e^{3/2} = 1.12$ is the Bondi accretion solution eigenvalue, and $u(x)$ is the velocity profile. In the inner regions, the Bondi solution is $u(x) \sim x^{-1/2}$, hence 
\begin{equation}
    \Gamma(x) = \frac{3}{2} \lambda c_s^3 \frac{\rho_\infty}{r_B} \frac{1}{x^3} \, . 
\end{equation}
Here, $c_s \simeq 15\ \rm km\ s^{-1}$ and $\rho_\infty = \mu m_p n_H \simeq (5-10) m_p\ \rm g\ cm^{-3}$ are the gas sound speed and density at infinity (molecular weight $\mu=1.22$ for a primordial gas composition), whose values are taken from the RHD profiles in Fig. \ref{fig:profiles}. 

The most important cooling process at $T=10^{4-5}$ K is Ly$\alpha$ collisional excitation. However, due to the very large H\,\textsc{i} column density of the system, Ly$\alpha$ photons are completely trapped, and the associated energy losses are negligible. In these conditions, we then assume that cooling is provided by free-free radiation, whose cooling rate [erg s$^{-1}$ cm$^{-3}$] is, recalling that for the Bondi solution $\rho(x) \sim x^{-3/2}$,      
\begin{equation}
    {\cal L} =  1.42\times 10^{-27} \sqrt{T} n_e n_p  = 1.42\times 10^{-27} n_\infty^2 x_e^2(T) \sqrt{T} \frac{1}{x^3} \, .
\end{equation}
where the ionization fraction is defined in terms of the electron or proton densities $x_e(T) = n_e /n_\infty \approx  n_p/n_\infty$. 

We first note that $\Gamma$ and ${\cal L}$ have the same radial dependence ($\sim x^{-3}$). This makes the temperature profile approximately constant. The gas temperature is obtained by equating the heating and cooling rates. With this procedure, we find that the equilibrium temperature is $T = 1.73 \times 10^6 n_\infty^{-2}$ K, or in the range $(1.7-6.9) \times 10^4$ K. Given the uncertainties and the approximations made, we adopt a fiducial value $T = 4 \times 10^4$ K when computing the emerging DCBH spectrum for LRDs.

\section*{Declarations}

\begin{itemize}
\item\textbf{Funding:} FP acknowledges support from the Black Hole Initiative at Harvard University. AF acknowledges support from the ERC Advanced Grant INTERSTELLAR H2020/740120. Partial support (AF) from the Carl Friedrich von Siemens Forschungspreis der Alexander von Humboldt-Stiftung Research Award is kindly acknowledged. We gratefully acknowledge the computational resources of the Center for High Performance Computing (CHPC) at SNS.
\item \textbf{Conflict of interest/Competing interests:} The authors declare no competing interests.
\item \textbf{Ethics approval and consent to participate:} Not applicable.
\item \textbf{Consent for publication:} Not applicable.
\item \textbf{Data availability:} 
The \textit{JWST} NIRSpec spectrum of LRD RUBIES-EGS~42046 is publicly available from the \href{https://mast.stsci.edu/portal/Mashup/Clients/Mast/Portal.html}{Mikulski Archive for Space Telescopes} at the Space Telescope Science Institute, which is operated by the Association of Universities for Research in Astronomy, Inc., under NASA contract NAS5-03127 for \textit{JWST}.
The synthetic spectra computed using the \textsc{GEMS} code are available from the authors upon request.
\item \textbf{Code availability:} The code developed for this study to match the spectrum of RUBIES-EGS~42046 is available from the authors upon request.
\item \textbf{Author contribution:} FP and AF presented the original idea in \citep{Pacucci_2015}, analyzed and interpreted the results; they led the writing of this paper. DDK provided the \textit{JWST} data used in this work and support for its analysis. All authors reviewed the manuscript.

\end{itemize}

\bibliography{ms}

\end{document}